\documentclass[aps,prl,twocolumn,groupedaddress,nofootinbib]{revtex4-2}  % for review and submission

\usepackage{graphicx}  % needed for figures
\usepackage{amsmath,amssymb}
\usepackage{siunitx}
\usepackage{flushend}
\usepackage{lipsum}
\usepackage{array}
\usepackage{float}
\usepackage{natbib}
\usepackage{booktabs}
\usepackage{color}
\usepackage{soul}
\usepackage{comment}

\begin{document}

\title{Origin of flat bands and non-trivial topology in coupled kagome lattices}

\author{Anumita Bose$^\S$}\email[]{anumitabose@iisc.ac.in}
\affiliation{Solid State and Structural Chemistry Unit,
Indian Institute of Science, Bangalore 560012, India.}

\author{Arka Bandyopadhyay$^\S$} 
\affiliation{Solid State and Structural Chemistry Unit, Indian Institute of Science, Bangalore 560012, India.}

\author{Awadhesh Narayan}\email[]{awadhesh@iisc.ac.in} \affiliation{Solid State and Structural Chemistry Unit, Indian Institute of Science, Bangalore 560012, India.}

\let\thefootnote\relax\footnote{$^\S$These authors contributed equally to this work.}

\vskip 0.25cm

\date{\today}

\begin{abstract}
We propose an exact analytical decimation transformation scheme to explore the fascinating coexistence of flat bands and Dirac fermions in three-dimensional coupled kagome systems. Our method allows coarse-graining of the parameter space that maps the original system to an equivalent low-level lattice. The decimated system enables defining a quantity in the tight-binding parameter space that predominantly controls the emergence of a flat band (FB) and provides a specific criterion for absolute flatness. Likewise, in terms of atomic separations, we develop a quantity that primarily controls the FB width in real materials and thus can be helpful in predicting new systems hosting FB as well as in tuning the FB width. Our predictions on the emergence of the flat band and Dirac fermions are confirmed for M$_3$X (M= Ni, Mn, Co, Fe; X= Al, Ga, In, Sn, Cr,...) family of materials, leveraging materials databases and first-principles calculations. Our work provides an analytical formalism that enables accurate predictions of FBs in real materials.
\end{abstract}

\maketitle

Flat bands (FBs), i.e., Bloch bands with vanishingly small dispersion, have spurred tremendous research interest in recent years due to their possibility to provide an ideal platform to study exotic many-body physics~\cite{yin2022topological}, such as ferromagnetism~\cite{sharpe2019emergent}, anomalous Landau levels~\cite{rhim2020quantum}, unconventional superconductivity~\cite{cao2018unconventional, balents2020superconductivity,peri2021fragile,volovik2018graphite}, non-Fermi liquid behavior~\cite{kumar2021flat}, and high temperature fractional quantum Hall effect~\cite{tang2011high, sheng2011fractional, neupert2011fractional,regnault2011fractional,mallik2023correlation}. In solid state systems, a FB with quenched kinetic energy can emerge due to two reasons -- isolated localized states without coupling to other states~\cite{kim2003direct,hu2018ubiquitous} or states resulting from a destructive quantum interference. Lattices with geometric frustration, such as kagome, Lieb, dice, and pyrochlore, fall into the latter category and generate localized eigenstates giving rise to FBs~\cite{sutherland1986localization,vidal1998aharonov,sun2011nearly,regnault2022catalogue,ma2020spin,jiang2019topological}.

In a kagome lattice, the single orbital nearest-neighbor hopping model provides real space eigenfunctions with opposite phases at neighboring sites~\cite{bergman2008band}. The destructive interference of these opposite phases prohibits any hopping to the adjacent cells, causing the confinement of the electronic state in the kagome hexagon. This, in turn, results in dispersionless bands in the momentum space. Furthermore, kagome lattices feature the existence of a Dirac point at the Brillouin zone (BZ) corner ($K$ point) and a saddle point at the zone boundary ($M$ point). However, the exact realization of these features in real materials has become challenging due to several other controlling factors, such as multiple orbital contributions, significant hopping beyond the nearest neighbor, and complex interlayer hopping. Recent studies have focused on the prediction as well as the experimental realization of FBs in the $k_{z}= 0$ plane, along with suppressed dispersion in other directions in the three-dimensional (3D) BZ in binary kagome metals TX (T: 3$d$ transition metal, X: Ge, Sn)~\cite{kang2020topological,kang2020dirac,liu2020orbital}. Other related systems such as Pd$_{3}$P$_{2}$S$_{8}$~\cite{park2020kagome} and YMn$_{6}$Sn$_{6}$~\cite{li2021dirac} exhibit spatially decoupled quasi-two-dimensional kagome planes. However, the impact of interlayer hopping in a 3D coupled kagome lattice is yet to be fully explored. To be more specific, we want to address the pertinent question: under what conditions do FBs and Dirac fermions coexist in these systems, and what are their energetic positions?

\begin{figure}[b]
\centering
\includegraphics[scale=0.10]{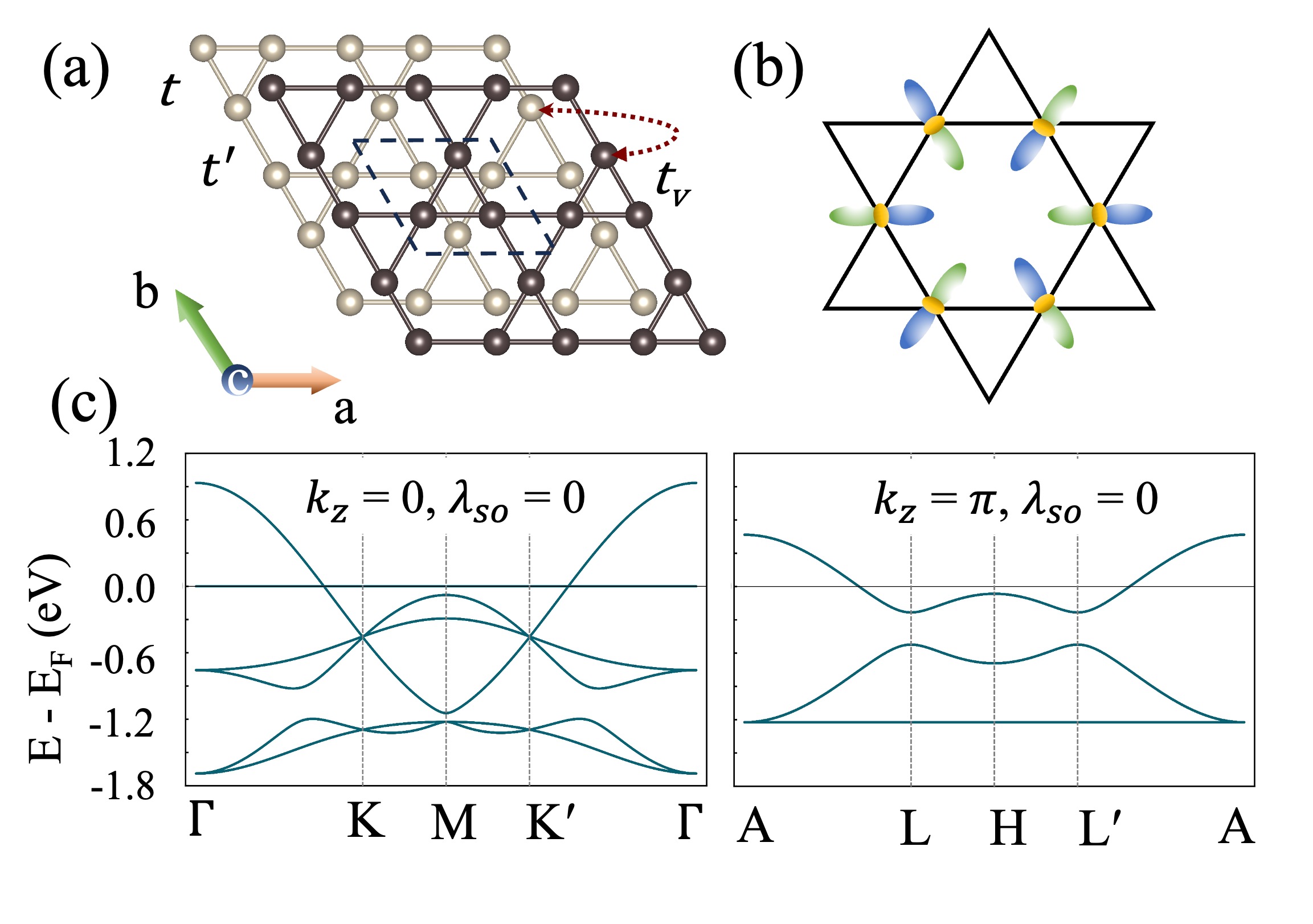}
  \caption{\textbf{Structure and electronic spectra for coupled kagome lattice model.} (a) Depiction of a coupled kagome lattice with intra- ($t$ and $t^{\prime}$) and inter- ($t_v$) layer hopping parameters. Two layers in the unit cell (marked by dashed lines) are denoted by different colours. (b) Schematic representation of the destructive interference for the $d_{xz}$ orbital in a kagome monolayer. (c) The emergence of a FB obtained from the TB model without the consideration of SOC for (left) $k_z$=0 plane; (right) $k_z$= $\pi$ plane. The FB in the $k_z = 0$ and $k_z = \pi$ planes reside at $E= \varepsilon+2t = 0$ eV and $E= \varepsilon -(t+t^{\prime}) = -1.22$ eV. Here, the parameters used are $t=0.33$ eV, $t^{\prime} = t/\sqrt{2}$, and $\varepsilon = -2t$.}
  \label{unitcell}
\end{figure}

In this work, we propose an exact analytical framework based on the real space decimation strategy to answer the above mentioned questions. In particular, the decimation scheme downfolds the original Hamiltonian without losing any information and provides the necessary conditions for a FB at $k_z=0$ plane and a Dirac point at $K$. Notably, the electronic spectra of the system in the $k_z=\pi$ plane resemble the monolayer kagome and, strikingly, follow the Su–Schrieffer–Heeger model~\cite{su1979solitons}. Incorporating spin-orbit coupling (SOC) in the model invariably lifts the band degeneracies, which can give rise to a nontrivial band topology. Our predictions have been confirmed by performing first-principles calculations on realistic candidate materials M$_3$X (M= Ni, Mn, Co, Fe; X= Al, Ga, In, Sn, Cr,...) obtained by leveraging materials database screening~\cite{jain2013commentary,vergniory2019complete}. Our scheme allows in predictions of FBs in a new material solely from its crystallographic information. This general recipe of flat band optimization also facilitates the engineering of a given FB by means of external perturbations, for example, pressure and strain. Our powerful approach, therefore, enables the exploration of the rich physics of such kagome systems.

\begin{figure}[t]
\centering
\includegraphics[scale=0.035]{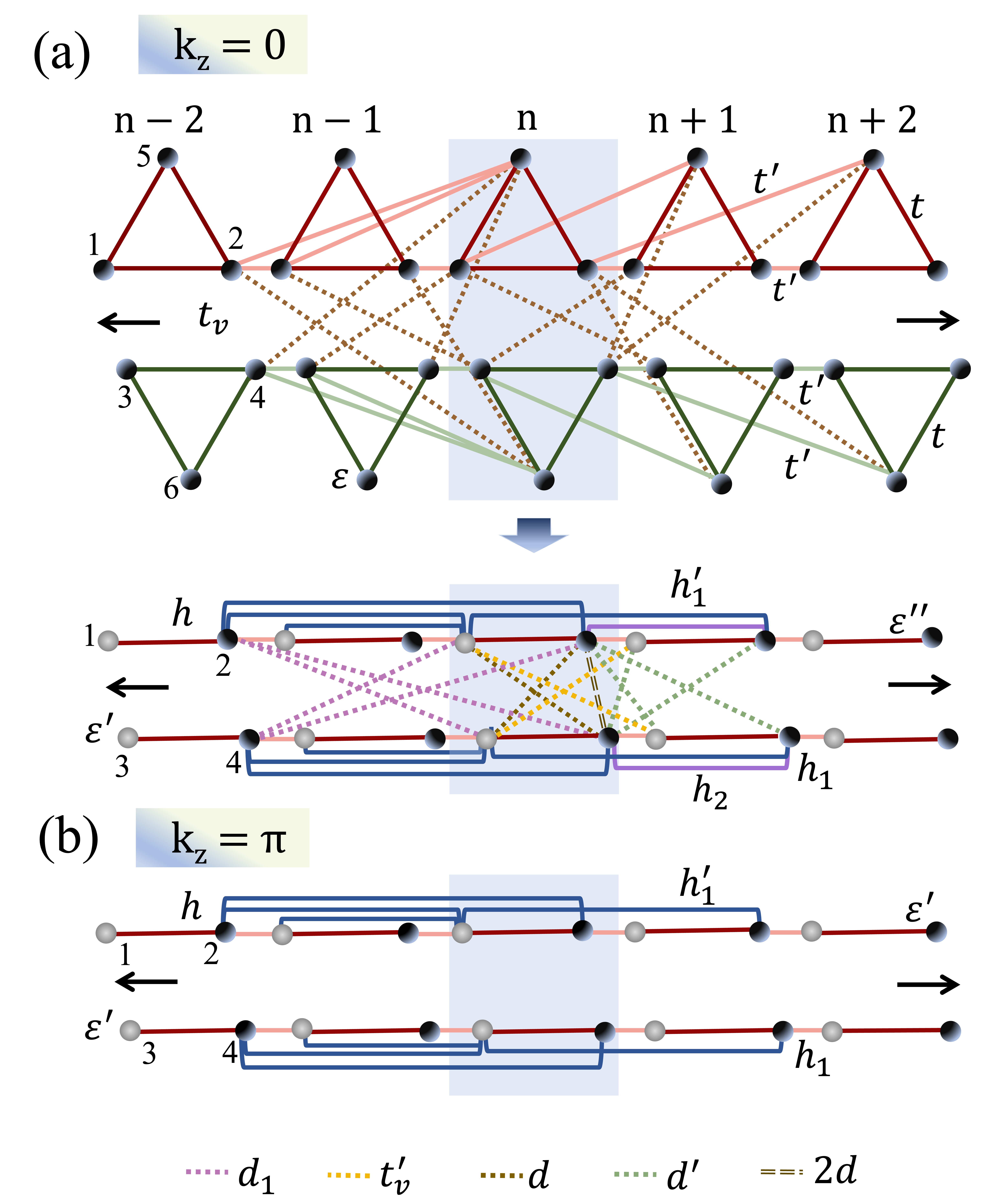}
  \caption{\textbf{Real space decimation scheme for coupled kagome lattice.} (a) Quasi-one-dimensional representation of the coupled kagome lattice keeping the Hamiltonian invariant along the symmetry path of the $k_{z}$ = 0 plane, i.e., $\Gamma-K-M-K^{\prime}-\Gamma$ (top). Here \{5, 6\} are decimated to get the equivalent downfolded four level ladder like network (bottom). (b) The ladder is decoupled into two identical linear chains of atoms for the $k_{z} = \pi$ plane. The shaded region indicates the unit cell with lattice parameter $a$/2. Other symmetric hopping terms are not shown for clarity.}
  \label{steps}
\end{figure}

\textit{Real space decimation method for coupled kagome system.} Our coupled kagome model consists of stacked kagome layers coupled by an inter-layer hopping $t_v$, along with two distinct intra-layer hopping parameters -- $t$ (intra-unit cell) and $t^{\prime}$ (inter-unit cell) as depicted in Fig.~\ref{unitcell}(a). One of the most intriguing properties of monolayer kagome lattice is the coexistence of Dirac fermions and localized states generated by the destructive interference of atomic orbitals [Fig.~\ref{unitcell}(b)]. However, the intra-layer hopping anisotropy and inter-layer coupling in coupled kagome lattices will drastically modify such an electronic spectrum. We have considered an effective quasi-one-dimensional (Q1D) model that precisely mimics the band structure of the original lattice in the $k_z=0$ plane, particularly along the high symmetry path $\Gamma-K-M-K^{\prime}-\Gamma$ [Fig.~\ref{unitcell}(c)]. We note that the choice of an equivalent Q1D lattice model is not very stringent in this case. For instance, Fig.~\ref{steps}(a) (top panel) depicts a typical example of a Q1D network for which the original tight-binding (TB) Hamiltonian $\hat{H} = \sum_i \varepsilon_i | i \rangle \langle i | + \sum_{i \neq j} t_{ij} | i \rangle \langle j | + H_{SO}$ remains invariant under the constraint $k_y=k_z=0$. In the above expression, $\varepsilon_i$ and $t_{ij}$ represent the onsite potential and hopping integral for the orbitals $|i \rangle$ and $|j \rangle$, respectively. The unit cell of the system contains six atomic sites that can be divided into the top and bottom layers denoted by site numbers \{1, 2, 5\} and \{3, 4, 6\}, respectively. Similar to the original lattice, there are two distinct intra-layer ($t$ and $t^{\prime}$) and one inter-layer ($t_v$) hopping parameters in this Q1D model. The SOC has been introduced by the term $H_{SO} = i \lambda_{SO} \sum_{\langle\langle i,j\rangle\rangle} \nu_{ij} c_{i}^{\dagger} s^{z} c_{j}$, where $s^{z}$ stands for the $z$ component of the Pauli matrix corresponding to spin, and $\nu_{ij} = \pm 1$ depends on the rotation direction while going from $j$ to the next-nearest neighbour site $i$~\cite{kane2005z}. In the momentum space, the Hamiltonian can be written in the orbital basis \{$| 1 \rangle , ... ,| 6 \rangle $\}, as follows

\begin{gather}
 H
 =
  \begin{bmatrix}
   \varepsilon & \Gamma & \Delta & 0 & \Theta & 0 \\
   \Gamma^* & \varepsilon & 0 & 0 & \Lambda & \Omega \\
   \Delta^* & 0 & \varepsilon & \Theta^* & 0 & \Gamma^* \\
   0   & 0 & \Theta & \varepsilon & \Xi & \Lambda \\
   \Theta^* & \Lambda^* & 0 & \Xi^* & \varepsilon & 0 \\
    0  & \Omega^* & \Gamma & \Lambda^* & 0 & \varepsilon 
   \end{bmatrix}.
   \label{eq:ham}
\end{gather}

In the above Eq.~\ref{eq:ham}, we have considered a uniform onsite potential $\varepsilon_i = \varepsilon \: (\forall i \in \{1, ..., 6\})$ while the other parameters are of the form, $\Gamma = t + t^{\prime} \: \exp[i(a_2-a_1)k_x]$, $\Delta = 2 t_v ( \exp[ia_2k_x] + \exp[i(a_2-a_1)k_x] )$, $\Theta = t + t^{\prime} \: \exp[ia_2k_x]$, $\Lambda = t + t^{\prime} \: \exp[ia_1k_x]$, $\Xi = 2 t_v ( \exp[ia_1k_x] + \exp[ia_2k_x])$, and $\Omega = 2 t_v ( \exp[ia_1k_x] + \exp[i(a_1-a_2)k_x] )$. The superscript $*$ denotes complex conjugation. 
We further have $a_1 = 2 a_2=a$ ($|\Vec{a}|=|\vec{b}|=a$). The transformation of this lattice model will be instrumental in unveiling the fundamental electronic structure of these coupled kagome lattices. For this purpose, we write down the TB analogue of the Schr\"odinger equation~\cite{chakrabarti1995role,sil2008metal,matulis2009analogy,bandyopadhyay2020review,bandyopadhyay20218} as $(E-\varepsilon_i)\psi_i - \sum_{i \neq j} t_{ij} \psi_j = 0$. The parameters, $E$, $\varepsilon_i$, $t_{ij}$ and $\psi_{i/j}$ stand for eigenenergy, onsite potential, hopping parameter, and wavefunction, respectively, while $i$ and $j$ are the site indices. This set of difference equations can be used to decimate a preferred subset of degrees of freedom or lattice sites -- this procedure maps the original system to its equivalent reduced version. In other words, a coarse-grained parameter space of TB Hamiltonian can be achieved using this exact formalism that does not lose any information about the original system. It is worth mentioning that the decimation process in various forms is the backbone of the real space renormalization group scheme~\cite{kadanoff1966scaling,refael2004entanglement,efrati2014real,agrawal2020quantum}. In the present case, we intend to absorb all the information of the lattice sites \{5, 6\} encoded in their probability amplitudes \{$\psi_5$, $\psi_6$\} into the remaining sites of the Q1D system. This decimation process can be performed using the following set of difference equations \\
\begin{align}
E^{\prime} \psi^{n}_{1} &= t \psi^{n}_{2} + t  \psi^{n}_{5}+t^{\prime} \psi^{n-1}_{2} +t^{\prime} \psi^{n+1}_{5}+t_{v} \psi^{n-1}_{3}+ t_{v} \psi^{n+1}_{3}, \nonumber \\
E^{\prime} \psi^{n}_{2} &= t \psi^{n}_{1} + t\psi^{n}_{5}  +t^{\prime} \psi^{n+1}_{1}   +t^{\prime} \psi^{n+2}_{5}   + t_{v} \psi^{n+1}_{6} + t_{v} \psi^{n+2}_{6}, \nonumber \\
E^{\prime} \psi^{n}_{5} & = t \psi^{n}_{1} + t  \psi^{n}_{2}  +t^{\prime} \psi^{n-2}_{2}   +t^{\prime} \psi^{n-1}_{1} + t_{v} \psi^{n-2}_{4} + t_{v} \psi^{n-1}_{4}.
\label{eq:diffexp}
\end{align}
In a similar vein, the difference equations for the bottom layer sites \{3, 4, 6\} can be written by considering the following transformations $\psi_1 \leftrightarrow \psi_3$, $\psi_2 \leftrightarrow \psi_4$, and $\psi_5 \leftrightarrow \psi_6$. In Eq.~\ref{eq:diffexp}, the parameter $E^{\prime}$ represents the potential energy contribution $E-\varepsilon$, and $n$ is the unit cell index. The decimation of the sites \{5, 6\} essentially downfolds the system to an exactly equivalent four-level lattice, shown in Fig.~\ref{steps}(a) (bottom panel). We note that this reduced system is obtained at the expense of allowing new energy-dependent onsite potentials and hopping terms that keep the characteristic equation invariant. In particular, the TB parameters of the renormalized lattice can be written in terms of the original ones as $\varepsilon^{\prime} = \varepsilon + (t^2 + {t^{\prime}}^{2})/(E-\varepsilon)$, $\varepsilon^{\prime \prime} = \varepsilon + (t^2+{t^{\prime}}^2+2t_{v}^{2})/(E-\varepsilon)$, $h_1=t^{\prime}+{t^{\prime}}^2/(E-\varepsilon)$, $h_{1}^{\prime}=t t^{\prime}/(E-\varepsilon)$, $h_2 = t_v^{2}/(E-\varepsilon)$, and $d=-t^{\prime}/(E-\varepsilon)$. 

The decimation process has redrawn the lattice problem to an equivalent depiction where two completely identical blocks made of lattice sites \{1, 2\} and \{3, 4\} with potential matrix $\Sigma (E,k_x)$ which are coupled with each other by the overlap matrix $T (E,k_x)$, which read

\begin{align}
 \Sigma &=     \begin{bmatrix}
   \varepsilon^{\prime} + h_{1}^{\prime} x_{p,n} &
  h + h_1 x_n + h_{1}^{\prime} x_{p,nn} \\
  h + h_1 x_p + h_{1}^{\prime} x_{n,pp} &
     \varepsilon^{\prime \prime} + h_{1}^{\prime} x_{pp,nn} + h_{2} x_{p,n}
   \end{bmatrix}, \nonumber \\ 
   T &=     \begin{bmatrix}
   t_v x_{p,n} &
 d+d_1 x_{nn} + d^{\prime} x_n \\
  d+d_1 x_{pp} + d^{\prime} x_p &
    2d + d^{\prime} x_{p,n} + d_1 x_{pp,nn}
   \end{bmatrix}. 
\label{eq:block}
\end{align}

Here $x_{\alpha,\beta}=x_\alpha+x_\beta$, where $x_\alpha$ and $x_\beta$ stand for the phase terms of the Bloch wavefunction. Further, $p$ and $pp$ represent the nearest neighbour and next-nearest neighbour inter-unit cell hopping contributions, respectively, in the positive $x$ direction, while the $n$ and $nn$ are the corresponding complex-conjugate terms. The Hermitian matrix $T (E,k_x)$ signifies that the dispersion relations of the decimated four level system will be identical to the eigenvalues of two $2 \times 2$ matrices given by $\Sigma (E,k_x) \pm T (E,k_x)$ [see, Supplementary information (SI) for detail expressions~\cite{xyz}]. Now, we focus on the above mentioned eigenvalues to explore the criteria for manifesting a non-dispersive band in the $k_z=0$ plane. It is fascinating that the condition $t_v = -t^{\prime}/2$ invariably eliminates all the dispersive terms from a particular eigenvalue, which gives rise to a FB in the $k_z=0$ plane at the energy value $E=\varepsilon + 2t$~\cite{xyz}. Here, it is worth mentioning that the presence of other interlayer hopping parameters viz. between neighbouring ($t_{v}^{\prime}$) and opposite vertices ($t_{v}^{\prime \prime}$) of the hexagon inside the unit cell [Fig.~\ref{unitcell}(a)] do not alter the condition for obtaining the FB. However, these parameters essentially shift the energetic position of the FB to $E=\varepsilon + 2 (t+ t_{v}^{\prime}) + t_{v}^{\prime \prime}$.

In addition, two bands touch each other at a Dirac point and sandwich another dispersive band at the $K$ point below the Fermi level for $t^{\prime}=t/\sqrt{2}$, as depicted in Fig.~\ref{unitcell}(c) (left panel). The inclusion of SOC ($\lambda_{so} \neq 0$) in the above case invariably lifts the band degeneracy (see SI ~\cite{xyz}) with a nontrivial topological index, i.e., $\mathbb{Z}_2=1$. Therefore, our lattice model not only reveals exciting FB physics but also exhibits the possibility of obtaining non-trivial topological phases. Furthermore, it can be shown from Eq.~\ref{eq:ham} that the Hamiltonian for $k_z=\pi$ plane transforms into that of the decoupled kagome lattices stacked along $c$-axis. Consequently, the band spectra resemble that of an isolated kagome lattice for $t^{\prime}=t/\sqrt{2}$, that possesses a FB away from the Fermi level at $E= \varepsilon -(t+t^{\prime})$ [Fig.~\ref{unitcell}(c) (right panel)]. In the $t=t^{\prime}$ case, two bands meet each other in the form of Dirac cones as typically obtained for pristine kagome lattice (see SI~\cite{xyz}). The set of difference equations given in Eq.~\ref{eq:diffexp} reveal that the above case is essentially the critical phase of the Su–Schrieffer–Heeger model~\cite{su1979solitons} and that the other two gapped phases ($t^{\prime} < t$ and $t^{\prime} > t$) are topologically distinct (see, SI for further details~\cite{xyz}). Nevertheless, quadratic band touchings are obtained between the flat and dispersive bands for all the cases in the $k_z=\pi$ plane, which are gapped upon inclusion of SOC, confirming its nontrivial topological nature~\cite{sun2011nearly}.\\

\begin{figure*}
\includegraphics[scale=0.11]{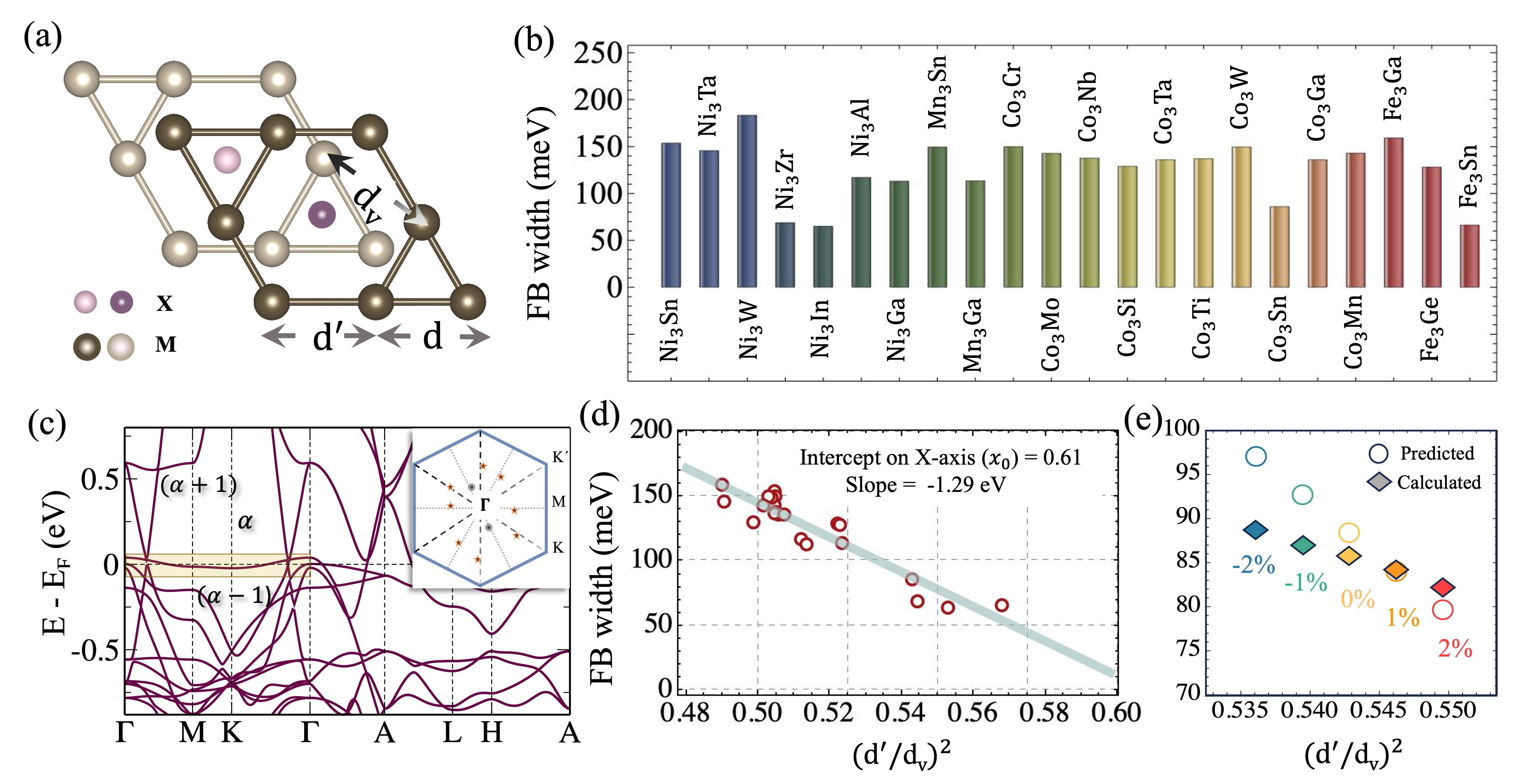}
  \caption{\textbf{Electronic structure and FBs in M$_3$X systems.} (a) Illustration of intra- ($d$ and $d^{\prime}$) and inter-layer ($d_v$) atomic separations in the $M_{3}X$ class of coupled kagome lattice. (b) A comparison of the DFT calculated FB width for M$_3$X systems. (c) Band structure for Ni$_3$In without the inclusion of SOC. The FB at the $k_z=0$ plane, indexed by $\alpha$, is marked by the yellow region. Inset shows the position of the Dirac points slightly away from the high symmetry directions on the $k_{z} = 0$ plane for Ni$_3$In. Gray circles and orange stars represent the Dirac points formed by the bands above and below the FB, respectively. (d) Variation of the FB width for the M$_3$X systems as a function of $(d'/d_{v})^2$. Here, each point represents a different material in the M$_3$X family. (e) FB engineering by controlling $(d'/d_{v})^2$ with the application of uniaxial strain (along $c$ axis) in Co$_3$Sn. Here, the positive (negative) value implies compressive (tensile) strain. The rhombus and circle represent the FB width obtained from DFT calculations and fitted line in Fig.~\ref{dft_bands}(d), respectively. }
  \label{dft_bands}
\end{figure*}

\textit{Density functional calculations for M$_3$X systems.} The above theory sets up a general formalism to reveal the underlying physics for FB formation in 3D coupled kagome lattices. Furthermore, we have traversed through the open-source materials databases -- Materials Project~\cite{jain2013commentary} and Topological Materials Database~\cite{vergniory2019complete} -- to corroborate our analytical findings for real materials. For that purpose, we have performed density functional theory (DFT) calculations to obtain the band structure of the screened systems (see SI for computational details~\cite{xyz}). Here, for a systematic analysis, we restrict our discussions to the materials belonging to the crystal class M${_3}$X [Space group P6$_3$/mmc (No. 194)] with easily recognizable coupled kagome geometry [Fig.~\ref{dft_bands}(a)] hosting a FB in the $k_z=0$ plane within the energy window of $\pm 1.2$ eV around the Fermi level. Moreover, the FB width has been calculated for all those systems as depicted in Fig.~\ref{dft_bands}(b). The lattice geometries of the candidate materials, their band structures, and the corresponding FB widths are presented in SI~\cite{xyz}.

Our calculations reveal that the band structure of the Ni$_3$In system, shown in Fig.~\ref{dft_bands}(c), exhibits the lowest FB width ($\sim$64.1 meV) among this class of materials. We note that Ni$_3$In  has recently been experimentally studied by Ye \textit{et al.}~\cite{ye2021flat}. A careful analysis indicates that despite the narrowest bandwidth, the FB in Ni$_3$In is slightly dispersive in nature with a maximum at the $\Gamma$ point and a minimum at the $K$ point. Dispersions of other systems also exhibit similar features. We have found similar FB dispersion in our TB model for the hopping relation $|t_v/t^{\prime}|<0.5$, as illustrated in SI~\cite{xyz}. Orbital analysis of Ni$_3$In reveals that the FBs at both the ($k_z=0$ and $\pi$) planes are mainly contributed by Ni $d$ orbitals -- namely $d_{xz}$/$d_{yz}$ orbitals (see SI~\cite{xyz}). The destructive quantum interference of Ni $d_{xz}$/$d_{yz}$ orbitals, as shown schematically in Fig.~\ref{unitcell}(b), results in the localization of electrons in a hexagonal region~\cite{ge2020d}. This electronic localization in the real space causes the Bloch wave functions to have a reduced dispersion in the momentum space, giving rise to the FB in this material.

It is worth mentioning that, in the TB model, we have used an effective single orbital basis per atom with different hopping, which is an idealized case and very difficult to realize in real systems. To give a more straightforward but equivalent flavor for the hopping parameters, we employ Harrison's universal inverse square scaling law~\cite{harrison}, which helps us to map the hopping parameters \{$t$, $t^{\prime}$, $t_v$\} to geometric parameters of the candidate systems, i.e., the separation between the atoms \{$d$, $d^{\prime}$, $d_v$\} [Fig.~\ref{dft_bands}(a)]. In particular, we have found that, ($d^{\prime}/d_{v})^2$, which is now proportional to $t_{v}/t^{\prime}$, plays the crucial role in controlling the FB width of these systems. We have plotted the FB width for the candidate materials with different ($d^{\prime}/d_{v})^2$ ratio, which exhibits a linear regression with a slope $\approx -1.29$ eV and an intercept on the horizontal axis, $x_{0}^{DFT} \approx 0.61$, as shown in Fig.~\ref{dft_bands}(d). The negative slope implies that the FB width decreases with increasing value of ($d^{\prime}/d_{v})^2$, while $x_0$ represents the condition for absolute flatness. Beyond this $x_0$ value of ($d^{\prime}/d_{v})^2$, the FB curvature is expected to change with the minimum and maximum at the $\Gamma$ and $K$ point, respectively, as supported by the TB model (see SI~\cite{xyz}). The variation of the FB width with $t_{v}/t^{\prime}$ obtained from the TB model also shows a similar linear variation with slope $\approx -1.29$ eV but $x_{0}^{TB}=0.5$, for $t=0.305$ eV and $t^{\prime}=1/\sqrt{2}$ (see SI~\cite{xyz}). Comparison between the $x_0$ values obtained from DFT and the TB model allows us to estimate the proportionality constant for ($d^{\prime}/d_{v})^2 \propto t_v/t^{\prime}$ to be 1.22. Moreover, the above value of the proportionality constant has been confirmed by comparing $(d^{\prime}/d_{v})^2 \approx 0.5528$ with the hopping ratio obtained by fitting the DFT results with the TB model $(t_v/t^{\prime})=0.45$ for Ni$_3$In. The above analysis reveals that the formation of FB in the coupled kagome systems is predominately controlled by the ratio between the intra- and inter-layer atomic spacing, i.e., $d^{\prime}/d_{v}$, which can help us in predicting the FB width of a new material in this class.

To establish the above-mentioned general recipe for optimizing the flatness, we have engineered the FB width with changing $d^{\prime}/d_{v}$ ratio, by applying strain. We have employed both compressive and tensile uniaxial strain along the $c$-axis for the Co$_3$Sn system, which essentially tunes the $d^{\prime}/d_{v}$ ratio towards and away from $x_{0}^{DFT}$, respectively. Thus, the compressive (tensile) strain, in turn, reduces (increases) the flatness as depicted in Fig.~\ref{dft_bands}(e).

\textit{Non-trivial topology of FB.} Next, we investigate the topological nature of the FB residing in the $k_z$ = 0 plane. We consider Ni$_3$In with a FB residing in the vicinity of the Fermi level. For clarity, we denote the FB with a band index $\alpha$, and neighbouring bands accordingly. First, we concentrate on the band structure in the absence of SOC. On both the sides, FB is touched by dispersive bands [the ($\alpha-1$)-th and ($\alpha+1$)-th bands] contributing to the Dirac points. The positions of these Dirac points are slightly away from the high symmetry directions on the $k_z$ = 0 plane and are shown in the inset of Fig.~\ref{dft_bands}(b). With the inclusion of SOC, these gapless points become gapped. Thus, we obtain a continuous direct band gap in the $k_z = 0$ plane between $\alpha$ and ($\alpha+1$) bands. Utilizing the inversion symmetry of the system, we use the Fu-Kane parity-based formulation~\cite{fu2007topological} to calculate the $\mathbb{Z}_2$ invariant in the $k_z$ = 0 plane. For Ni$_3$In, we obtain $\mathbb{Z}_2=1$, confirming the non-triviality of the gap between the $\alpha$-th and $(\alpha+1)$-th bands in the $k_z =0$ plane. We have also found a similar non-trivial behaviour in other materials of this family -- Ni$_3$Al and Ni$_3$Ga. As we discussed earlier, our TB analysis with the hopping relations obtained from the decimation method supports the non-trivial topological phase of these 3D coupled kagome systems in presence of SOC.
In a nutshell, we present a decimation transformation technique to unravel the reappearance of the FB and Dirac cones in 3D coupled kagome lattices. This method essentially downfolds the tight-binding Hamiltonian by decimating some preferred lattice sites at the expense of energy-dependent renormalized onsite and hopping parameters. The equivalent low-level system, in turn, reveals that the interlayer hopping is crucial in generating a FB in the $k_z=0$ plane. Furthermore, incorporating SOC in the model invariably triggers a nontrivial topological phase characterized by $\mathbb{Z}_2 = 1$. Our predictions on the emergence of the FB and Dirac fermions are confirmed by first-principles calculations for a class of kagome materials. We find our approach helpful in predicting the possibility of hosting a FB in coupled kagome lattice by simply analysing the lattice geometry. Our theory also suggests the engineering of FB width by applying external perturbation, e.g., strain and pressure. The present work allows exploration of the underlying physics of the intricate electronic properties of the coupled kagome lattices and can further assist in designing other lattice systems with nontrivial topological FBs and Dirac fermions.

\subsection{Acknowledgements:} A. Bose acknowledges Prime Minister's Research Fellowship. A. Bandyopadhyay thanks the Indian Institute of Science IoE postdoctoral fellowship for financial support. A.N. acknowledges support from the start-up grant (SG/MHRD-19-0001) of the Indian Institute of Science.

%\def\thefootnote{\bot}\footnotetext{These authors contributed equally to this work}\def\thefootnote{\arabic{footnote}}
%$^\bot$ These authors contributed equally to this work. 

\bibliographystyle{apsrev}
\bibliography{ref}

\onecolumngrid

\clearpage

\section{Supplementary Information}

\section{Band dispersion relations for decimated model}

The band dispersion relations of the four-level system obtained through the decimation scheme given in Fig. 2 of the main manuscript along the path $k_z=0$ plane can be written as follows. \\

\begin{align}
    \lambda_1 & = \frac{1}{2} \left( (\xi_1 + \xi_2) + \sqrt{(\xi_1 + \xi_2)^2 - 4 (\zeta_1 + \zeta_2)}\right)  \nonumber \\
    \lambda_1 & = \frac{1}{2} \left( (\xi_1 + \xi_2) - \sqrt{(\xi_1 + \xi_2)^2 - 4 (\zeta_1 + \zeta_2)}\right) \nonumber \\
        \lambda_3 & = \frac{1}{2} \left( (\xi_1 - \xi_2) + \sqrt{(\xi_1 - \xi_2)^2 - 4 (\zeta_1 - \zeta_2)}\right) \nonumber \\
    \lambda_4 & = \frac{1}{2} \left( (\xi_1 - \xi_2) - \sqrt{(\xi_1 - \xi_2)^2 - 4 (\zeta_1 - \zeta_2)}\right) 
    \label{eq:foureig}
\end{align}

In the above eq.~\ref{eq:foureig}, the parameters depend on the eigenenergy and momentum that make the characteristic equation invariant upon decimation. The parameters can be explicitly written down as given below.   \\

$\xi_1 = \varepsilon^{\prime} + \varepsilon^{\prime \prime} + h_{1}^{\prime} \; {x_{n}} + h_{2} \; {x_{n}} + h_{1}^{\prime} \; x_{nn} + h_{1}^{\prime} \;{x_{p}}+ h_{2} \;{x_{p}}+ h_{1}^{\prime} \; x_{pp}$, \\

$\xi_2 = 2 d + d \; {x_{n}} + {d_{1}} \; {x_{n}} + t_{v} \; {x_{n}} + {d_{1}} \; x_{nn} + d \;{x_{p}}+ {d_{1}} \;{x_{p}}+ t_{v} \;{x_{p}}+ {d_{1}} \;x_p$, \\

$\zeta_1 = \varepsilon^{\prime} \; \varepsilon^{\prime \prime} - h^2 + \varepsilon^{\prime \prime} \; h_{1}^{\prime} \; {x_{n}} + \varepsilon^{\prime} \;h_{2} \; {x_{n}} - h_{1} \;h_{1}^{\prime} \; {x_{n}}^2 + h_{1}^{\prime} \; h_{2} \; {x_{n}}^2 +
  {d_{1}} \; t_{v} \; {x_{n}}^2 + \varepsilon^{\prime} \; h_{1}^{\prime} \; x_{nn} + {d_{1}} \; t_{v} \; {x_{n}} \;x_{nn} + \varepsilon^{\prime \prime} \; h_{1}^{\prime} \;{x_{p}}+ \varepsilon^{\prime} \; h_{2} \;{x_{p}}- 
 {d_{1}}^2 \; {x_{n}} \;{x_{p}}- h_{1}^2 \; {x_{n}} \;{x_{p}}- {h_{1}^{\prime}}^2 \; {x_{n}} \;{x_{p}}+ 2 \; h_{1}^{\prime}\; h_{2}\; {x_{n}}\;{x_{p}}+ 
 2 \; {d_{1}} \; t_{v} \; {x_{n}} \;{x_{p}}- {d_{1}}^2 \; x_{nn} \;{x_{p}}- h_{1} \; h_{1}^{\prime} \; x_{nn} \;{x_{p}}+ {h_{1}^{\prime}}^2 \; x_{nn} \;{x_{p}}+ 
 {d_{1}} \; t_{v} \; x_{nn} \;{x_{p}}- h_{1} \; h_{1}^{\prime} \;{x_{p}}^2 + h_{1}^{\prime} \; h_{2} {x_{p}}^2 + {d_{1}} \; t_{v} \; {x_{p}}^2 - 
 d^2 \; (1 + {x_{n}}) (1 + {x_{p}}) + 
 d \; t_{v} ({x_{n}} + {x_{p}}) (2 + {x_{n}} + {x_{p}}) + (\varepsilon^{\prime} \; h_{1}^{\prime} - {d_{1}}^2 ({x_{n}} + x_{nn}) + 
    h_{1}^{\prime} (-h_{1} \; {x_{n}} + h_{1}^{\prime} \; {x_{n}} - h_{1}^{\prime} \; x_{nn}) + {d_{1}} \; t_{v} ({x_{n}} + {x_{p}})) x_{pp} - 
 h (h_{1} ({x_{n}} + {x_{p}}) + h_{1}^{\prime} ({x_{n}} + x_{nn} +{x_{p}}+ x_{pp})) - 
 d {d_{1}} (x_{nn} +{x_{p}}+ x_{nn} \;{x_{p}}+ x_{pp} + {x_{n}} (1 + 2{x_{p}}+ x_{pp}))$, \\
 
 $\zeta_2=-d (h_{1} \; {x_{n}} - h_{1}^{\prime} \; {x_{n}} + h_{1}^{\prime} \; x_{nn} + h_{1} \;{x_{p}}- h_{1}^{\prime} \;{x_{p}}+ 2 \; h_{1} \; {x_{n}} \;{x_{p}}- 
    2 \; h_{1}^{\prime} \; {x_{n}} \;{x_{p}}+ h_{1}^{\prime} \; x_{nn} \;{x_{p}}- \varepsilon^{\prime} (2 + {x_{n}} + {x_{p}}) + h (2 + {x_{n}} + {x_{p}}) + 
    h_{1}^{\prime} (1 + {x_{n}}) x_{pp}) + 
 t_{v} ({x_{n}} + {x_{p}}) (\varepsilon^{\prime \prime} + h_{2} ({x_{n}} + {x_{p}}) + h_{1}^{\prime} (x_{nn} + x_{pp})) + 
 {d_{1}} (-2 h_{1} \; {x_{n}} \;{x_{p}}+ 2 h_{1}^{\prime} \; {x_{n}} \;{x_{p}}- h_{1} \; x_{nn} \;{x_{p}}- h_{1} \; {x_{n}} \; x_{pp} - 
    2 \; h_{1}^{\prime} \; x_{nn} \; x_{pp} + \varepsilon^{\prime} ({x_{n}} + x_{nn} +{x_{p}}+ x_{pp}) - 
    h ({x_{n}} + x_{nn} +{x_{p}}+ x_{pp}))$. \\

In the above expressions, we have considered $x_{p} = \exp[i k_x a/2]$, ${x_{n}} = \exp[-i k_x a/2]$, $x_{pp} = \exp[i k_x a]$, and $x_{nn} = \exp[i k_x a]$. We now focus on the eigenvalue $\lambda_1$ to explore the criteria for manifesting a non-dispersive band in the $k_z=0$ plane. It is fascinating that the condition $t_v = -t^{\prime}/2$ substantially simplifies the expressions of of $(\xi_1 + \xi_2)$ and $(\zeta_1 + \zeta_2)$ as $2 \varepsilon^{\prime} + 4 t_v \textup{cos}(k/2)$ and ${\varepsilon^{\prime}}^{2} - (t^{\prime} + 2 d_1)^2 - (d+h)^2 + 4 t_v (\varepsilon - t) \textup{cos}(k/2)$, respectively. Substituting these values in the eigenvalue mentioned above, remarkably, eliminates all the dispersive terms giving rise to a flat band in the $k_z=0$ plane at the energy value $E=\varepsilon + 2t$.

\section{Methods}

 We have performed the electronic structure calculations by using density functional theory, as encoded in the
QUANTUM ESPRESSO code ~\cite{giannozzi2009quantum,giannozzi2017advanced}. We have employed generalized gradient approximation based on Perdew-BurkeErnzerhof parametrization~\cite{perdew1996generalized} within projector augmented
wave basis ~\cite{blochl1994projector}. Band structures were calculated both with and without the inclusion of spin-orbit interaction (SOI) using scalar and fully relativistic pseudopotentials, respectively. A cutoff value of 75 Ry was chosen for plane waves. A 12 × 12 × 15 Monkhorst-Pack grid ~\cite{monkhorst1976special} is used for the self-consistent calculation. After the self-consistent calculation, maximally localized
Wannier functions were constructed using the WANNIER90 package ~\cite{mostofi2008wannier90} considering Ni 3d and In 3p orbitals as the basis. The positions of the Dirac points on the k$_z$ = 0 plane were calculated using the Nelder and Mead’s Downhill Simplex Method~\cite{nelder1965mead}, as implemented in WannierTools~\cite{wu2018wanniertools}, which takes a uniform $k$-mesh in the three-dimensional Brillouin zone as a set of starting points for the calculation. 

The 2D hexagonal plane (k$_{z}$=0) of the 3D BZ has four time-reversal invariant momenta (TRIM) points --- $\Gamma$ and three $M$ points. Under the inversion symmetry of the system, we use the Fu-Kane parity-based formulation to calculate $\mathbb{Z}_2$ invariant~\cite{fu2007topological}. The $\mathbb{Z}_2$ invariant for the $k_{z}=0$ plane $\nu_{k_{z}=0}$ can be defined using $(-1)^{\nu_{k_{z}=0}}=\Pi_{i=1}^\alpha \Pi_{\Gamma_{j}}   \chi_{2i}(\Gamma_{j})$, where `$i$' runs over the occupied bands and $\Gamma_j$ represents a TRIM point on the $k_{z}=0$ plane. $\chi_{2i}$ $(\Gamma_{j}$) denotes the parity eigenvalue for the `$2i$' th band at TRIM $\Gamma_{j}$, and can take value $\pm$1.\\

\section{Crystal structure and electronic properties of M$_3$X systems}

\subsection{Crystal structure:}
M$_3$X (M = Ni, Mn, Co; X=Al, Ga, In, W, Sn, Nb, Si, Ge, Mo) crystallizes in the layered hexagonal lattice with non-symmorphic space group (SG) P63/mmc (No. 194). The unit
cell consists of two layers of atoms. Each layer contains four atoms – three `M' atoms and one `X' atom. Atoms of the two such layers stacked along the `c' direction are connected by inversion symmetry. 

\begin{figure}[H]
\centering
\includegraphics[scale=0.1]{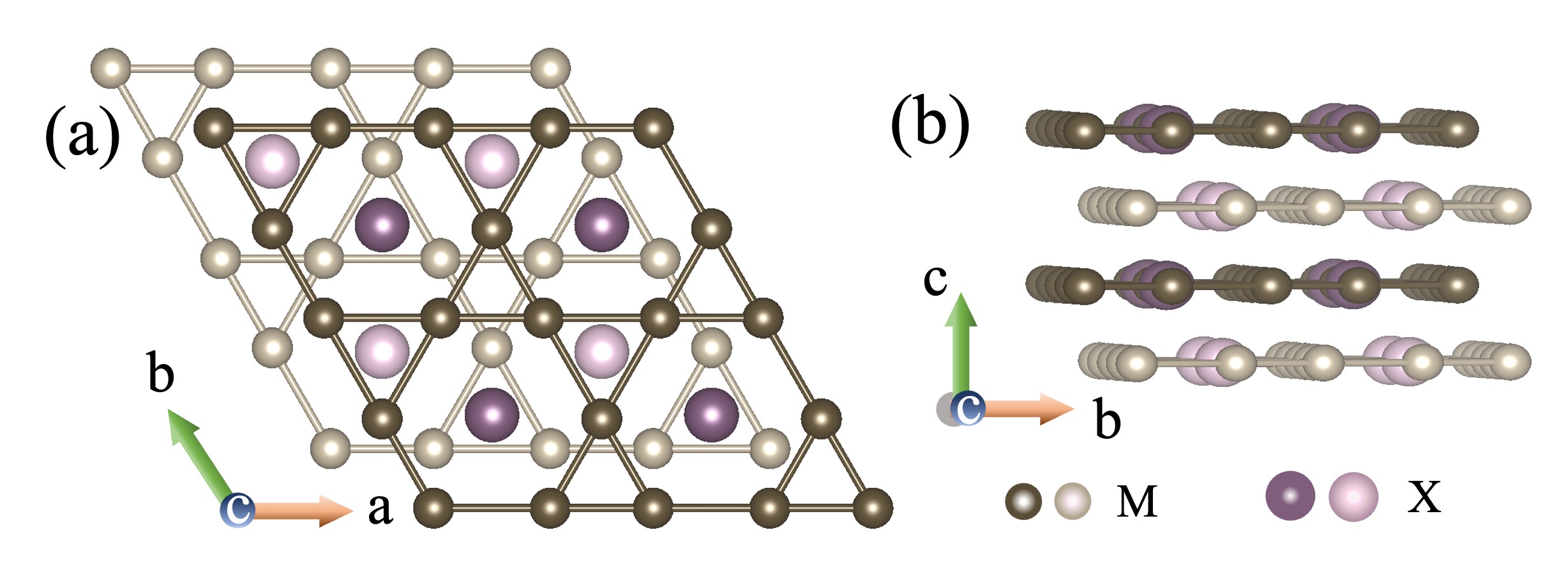}
   \caption{\textbf{Structure of the M$_{3}$X systems:} (a) top view, (b) side view. Dark and light colored spheres denote the top and bottom layers, respectively.}     \label{structure}
\end{figure}

Fig.~\ref{structure} presents the (a) top view and (b) side view of a coupled kagome lattice. Three `M' atoms in a layer form a Kagome-type lattice – with triangles and hexagons, and the remaining `X' atom is positioned at the center of the hexagon. 
In this work, we have highlighted twenty-three systems of the coupled kagome family with the same space group, P63/mmc, obtained through materials search~\cite{jain2013commentary,vergniory2019complete}. The lattice geometry in terms of the atomic separations -- d, d$^\prime$, and d$_v$ [see, main manuscript: Fig~3(a)] of these systems have been enlisted in Table.~\ref{table: width}. \\

\subsection{Band dispersion for M$_3$X systems:}
\begin{figure}[h!]
\begin{center}
\includegraphics[height=20cm,width=16cm]{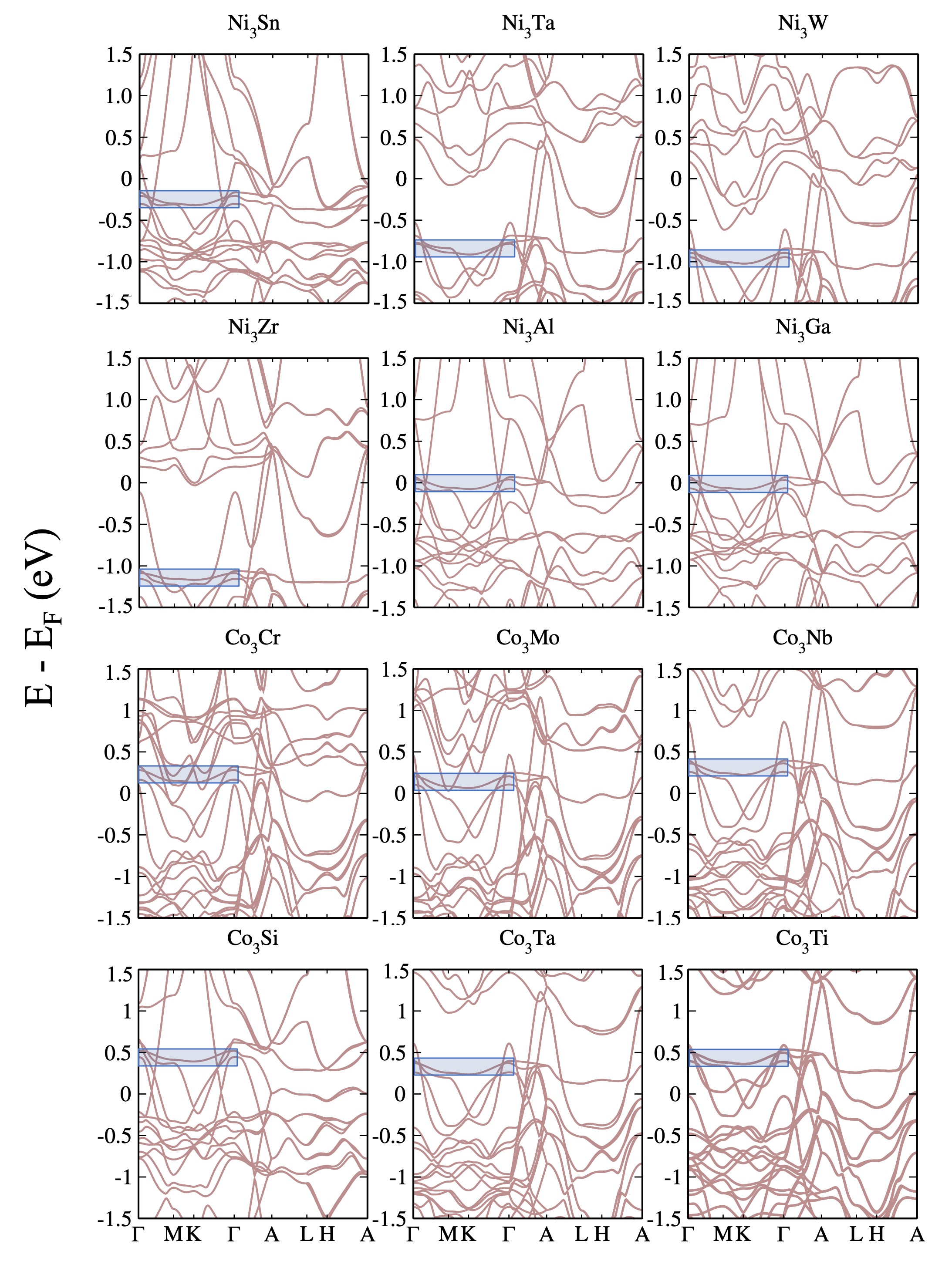}
\caption{\textbf{Electronic structures for the M$_3$X systems using the density functional theory} M$_3$X systems host a flat band in the $k_z$ = 0 plane, highlighted in the blue shaded box.}   \label{dft_M3X_bands1}
\end{center}
\end{figure}

\begin{figure*}[h!]
\begin{center}
\includegraphics[height=16cm,width=19cm]{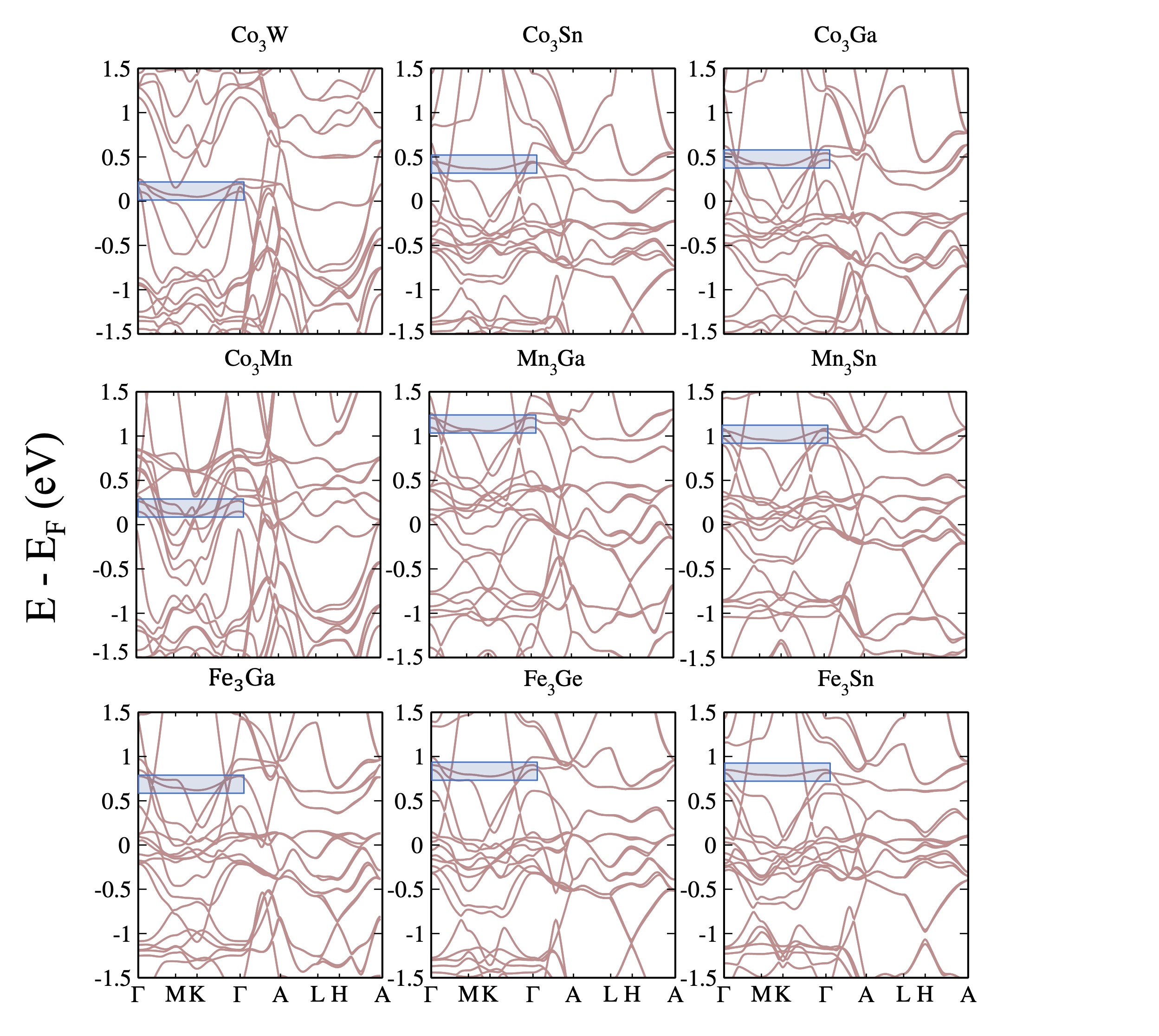}
\caption{\textbf{Electronic structures for the M$_3$X systems using the density functional theory} M$_3$X systems host a flat band in the $k_z$ = 0 plane, highlighted in the blue shaded box.}   \label{dft_M3X_bands2}
\end{center}
\end{figure*}

In the main manuscript, we have discussed the band dispersion for the Ni$_3$In system [see, main manuscript: Fig~3(c)]. Here in Fig.~\ref{dft_M3X_bands1} and ~\ref{dft_M3X_bands2}, we present the electronic band dispersions in presence of spin orbit interaction for the rest of the systems. All these systems host a flat band in the k$_z$ = 0 plane. The corresponding width of the flat bands (presented in Fig. 3(b) of the main text) have been consolidated in Table.~\ref{table:width}. \\

\begin{table}[h!]
\caption{\textcolor{black}{Structures and FB width in M$_3$X systems.}}
\label{table:width}
\begin{tabular}{|p{2.4cm}|p{2.4cm}|p{2.4cm}|p{2.4cm}|p{2.4cm}|p{2.4cm}|}
 \toprule
 %\multicolumn{4}{|c|}{Unit cell information} \\
  System & d (\AA)  & d$^\prime$ (\AA)& d$_v$ (\AA)& (d$^\prime$/d$_v$)$^2$ & FB width (meV)\\
 \hline
 \hline
Ni$_3$Sn  &2.6475&2.6385&3.7147&0.5045&153.7\\

Ni$_3$Ta 
&2.5650& 2.5650&3.6626&0.4904&145.6\\

Ni$_3$W  
&2.5450&2.5500&3.6114&0.4986&130.2\\

Ni$_3$Zr  
&2.4618&2.8028&3.799&0.5443&68.8\\

Ni$_3$In 
&2.5104& 2.8421&3.8226&0.5528&64.1\\

Ni$_3$Al 
&2.5063& 2.5530&3.5684&0.5118&117.1\\

Ni$_3$Ga 
&2.5119& 2.5750&3.5939&0.5134&113.0\\

Mn$_3$Ga 
&2.7046&2.6594&3.8012&0.5028&149.5 \\

Mn$_3$Sn  
&2.7162&2.8749&3.9740&0.5233&113.5\\

Co$_3$Cr  &2.5183&2.5097&3.5328&0.5047&150.0\\ 

Co$_3$Mo 
&2.5694&2.5606&3.6051&0.5045&142.8\\

Co$_3$Nb  &2.5815&2.5815&3.6337&0.5047&137.6\\        
Co$_3$Si
&2.4027&2.5733&3.5617&0.5220&129.0\\

Co$_3$Ta  
&2.5855&2.5855&3.6364&0.5056&136.0\\ 

Co$_3$Ti  &2.5595&2.5615&3.6063&0.5045&136.9\\

Co$_3$W 
&2.5644&2.5556&3.6010&0.5037&149.4\\        
Co$_3$Sn  
&2.5213&2.7982&3.7980&0.5428&85.9\\

Co$_3$Ga  
&2.5280&2.5577&3.5911&0.5072&135.9\\

Co$_3$Mn
&2.4829&2.4475&3.4565&0.5014&143\\

Fe$_3$Ga  &2.6044&2.5956&3.7082&0.4899&159.2\\

Fe$_3$Ge  
&2.5004&2.6780&3.7041&0.5227&128.1\\

Fe$_3$Sn  
&2.4717&2.9867&3.9635&0.5678&66.3\\
\hline
\end{tabular}
\end{table}

\newpage

\newpage

\subsection{Orbital projected bands for the Ni$_3$In system:}
\begin{figure*}[h!]
  \includegraphics[scale=0.2]{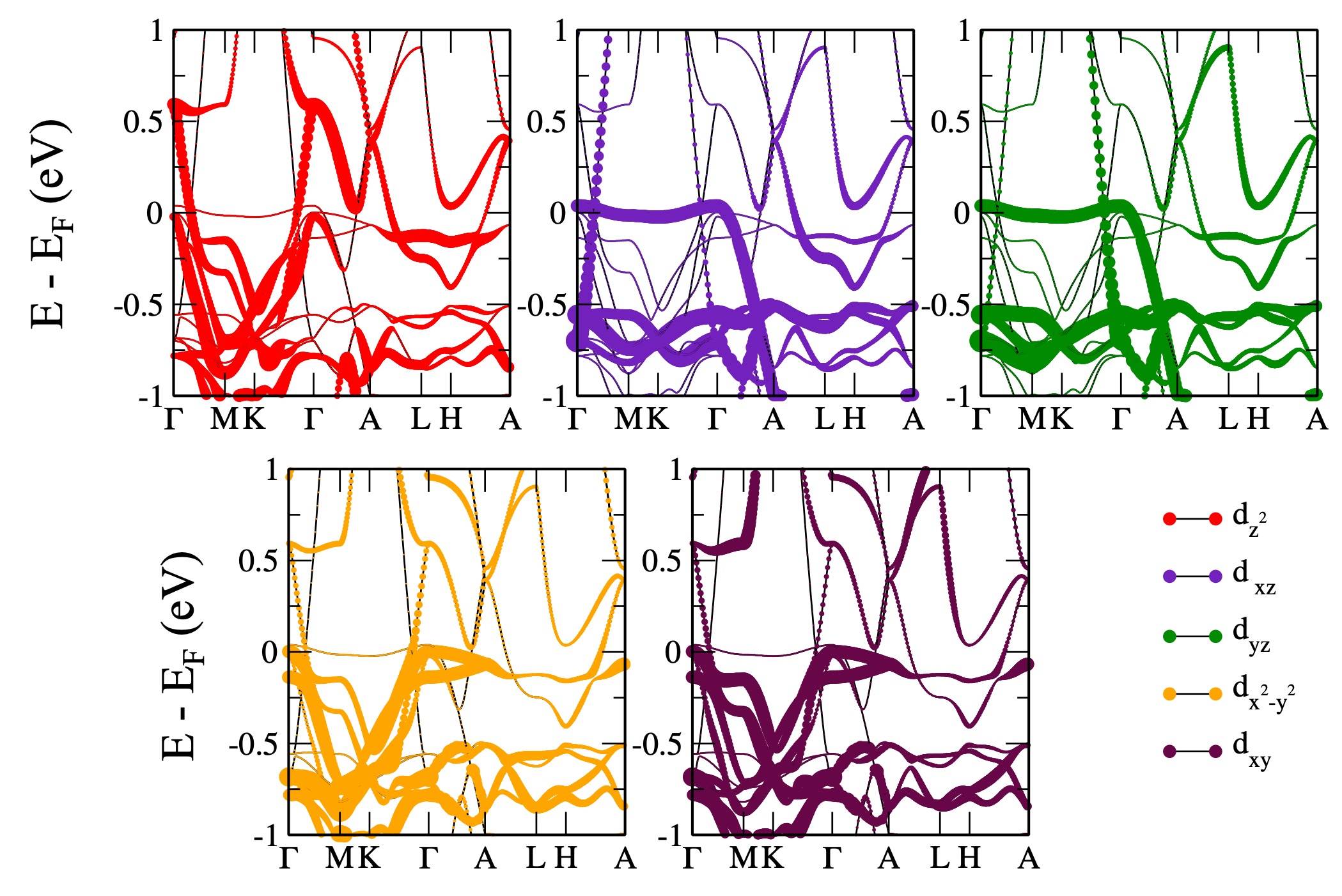}
   \caption{\textbf{Orbital resolved bands:} Orbital resolved electronic spectra for the Ni d orbitals in the Ni$_3$In system shows the dominance of d$_{xz}$ and d$_{yz}$ orbitals in the flat bands for both $k_z = 0$ and $k_z = \pi$ planes. }    \label{fatbands}
\end{figure*}

In Fig.~\ref{fatbands} we present the Ni-d orbital projected bands for the Ni$_3$In system system. We observe that the flat bands at both the k$_z$ = 0 and k$_{z} = \pi$ planes are majorly contributed by d$_{xz}$ and d$_{yz}$ orbitals.

\section{Effects of Spin Orbit Coupling:}

\begin{figure*}[h!]
\includegraphics[scale=0.15]{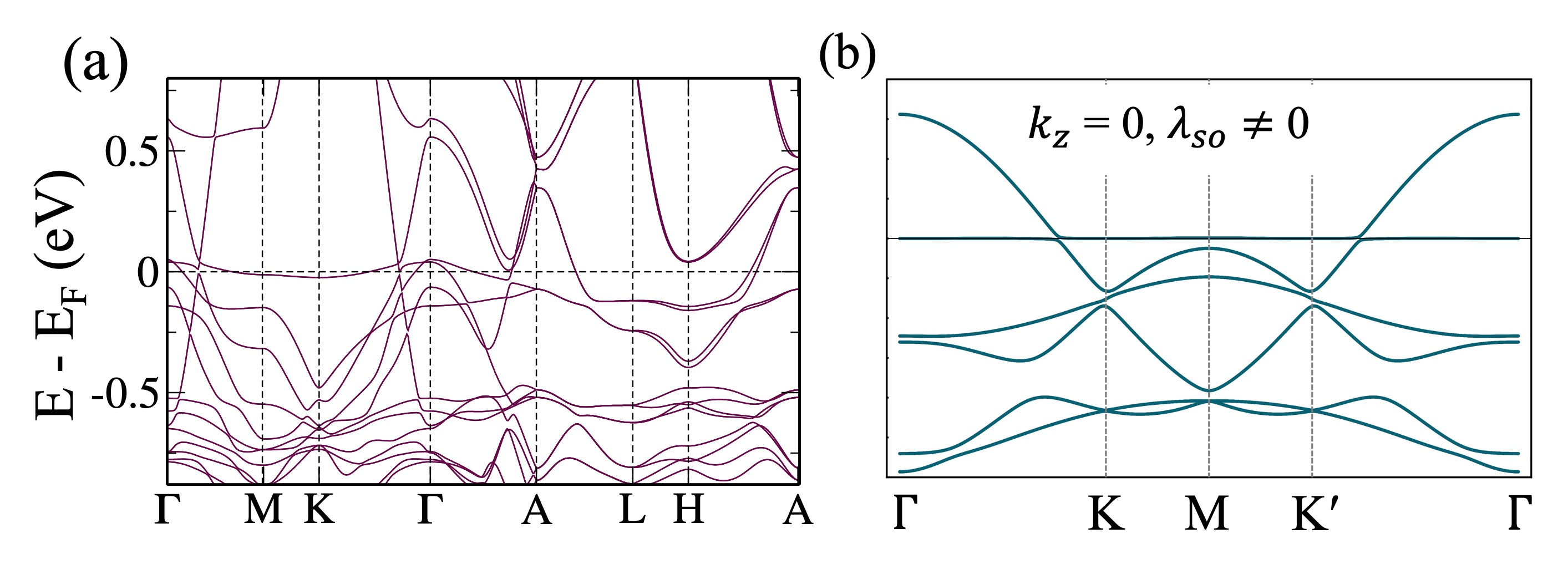}
\caption{\textbf{Band dispersion in coupled kagome systems with SOC: }Bandstructures with the inclusion of SOC for coupled kagome systems, obtained from (a) DFT calculations (for Ni$_3$In) and (b) TB model ($\lambda_{so} = 0.06 \: t$). Effects of SOC make each band doubly degenerate. On the $k_{z} = 0$ plane, except for the gap opening at the Dirac points, the bands look almost similar to their `without SOC' counterparts, , respectively, as presented in FIGs 3(c) and 1(c) (left panel) of the main manuscript.}  \label{soc}
\end{figure*}

In the main manuscript, we have presented band structures calculated from the TB model [see, main manuscript: Fig. 1(c)] and DFT computations [for Ni$_3$In, see, main manuscript: Fig. 3(c)] without incorporating SOC. Here, we calculate the dispersions in the presence of the spin-orbit coupling (SOC). In Fig.~\ref{soc} (a) and (b), we have depicted the band dispersions obtained from DFT calculations and TB model, respectively. We note that, for Ni$_3$In, the intrinsic SOC value is small. Our TB model considers the SOC strength to be $\lambda_{so} = 0.06 \:t$. In both cases, some band degeneracies, including the Dirac point at $K$ point are lifted upon incorporating SOC, which can readily be seen in FIG.~\ref{soc}.

\section{Hopping parameter dependent band dispersions:}
\subsection{For $k_z = 0$ plane}
\begin{figure*}[h]
\includegraphics[scale=0.12]{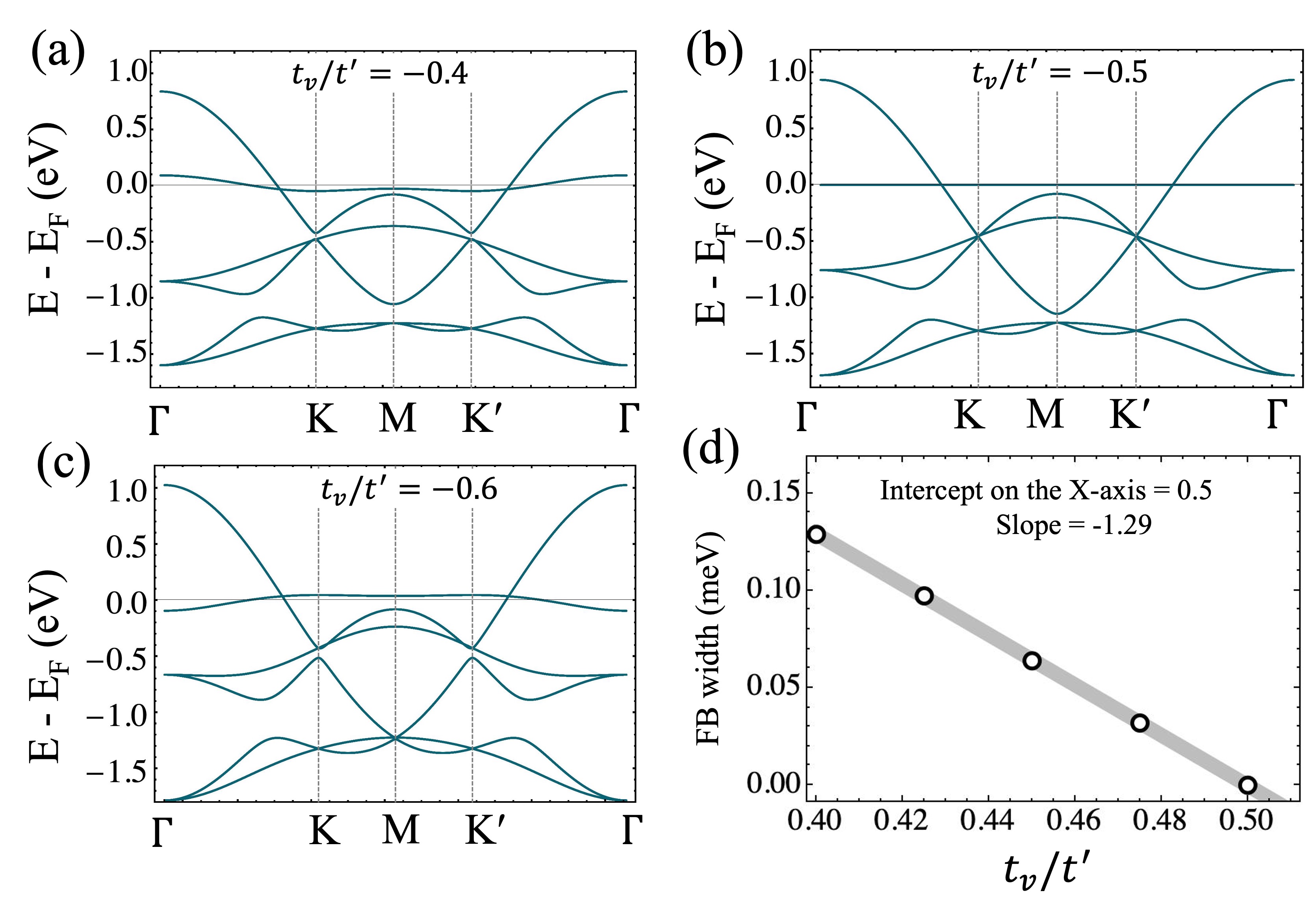}
\caption{\textbf{Band dispersions obtained from the TB model for different values of the intra-layer hopping parameters t and t$^\prime$ for the $k_z = 0$ plane.} An absolute flat band at the Fermi level and a Dirac point at $K$ appear for $t_{v}/t^{\prime} = -0.5$ (in (b)). Values greater (in (a)) or smaller (in (c)) than `-0.5' make the flat band slightly dispersive with curvature opposite to each other. (d) shows the variation of FB width as a function of $t_{v}/t^{\prime}$.}   \label{tv_tp_bands}
\end{figure*}

In the maintext, we have explained the condition to achieve the absolute flatness from the TB model, which reads $t_{v}/t^{\prime} = -0.5$. In Fig.~\ref{tv_tp_bands}, we try to find the band dispersions for a variation for the above-mentioned ratio. The middle panel presents the dispersion for $t_{v}/t^{\prime} = -0.5$, with an absolute flat band at the Fermi energy and a Dirac point at $K$. We observe a slight variation of $t_{v}/t^{\prime}$ from -0.5 makes the flat band slightly dispersive. Also, the bands forming Dirac point at $K$ now become gapped. The left panel depicts the bandstructure $t_{v}/t^{\prime} = -0.4$. The slightly dispersive band has a minimum at $K$ and a maximum at $\Gamma$. Flat band for $t_{v}/t^{\prime} = -0.6$ features the opposite nature with a maximum at $K$ and a minimum at $\Gamma$, as presented in the right panel. It is worth mentioning that for all the systems we have studied, a flat-like band follows the curvature for that of the $|t_{v}/t^{\prime}| < 0.5$ case. The dispersive bands are also gapped exactly at $K$ point. Although they could be present slightly away from $K$. The exact position for the Dirac point in Ni$_3$In system was found to be (0,0,0.07), i.e., slightly shifted along the $k_z$ direction from the $\Gamma$ point.

\subsection{For $k_z = \pi$ plane}

\begin{figure*}[h!]
\includegraphics[scale=0.115]{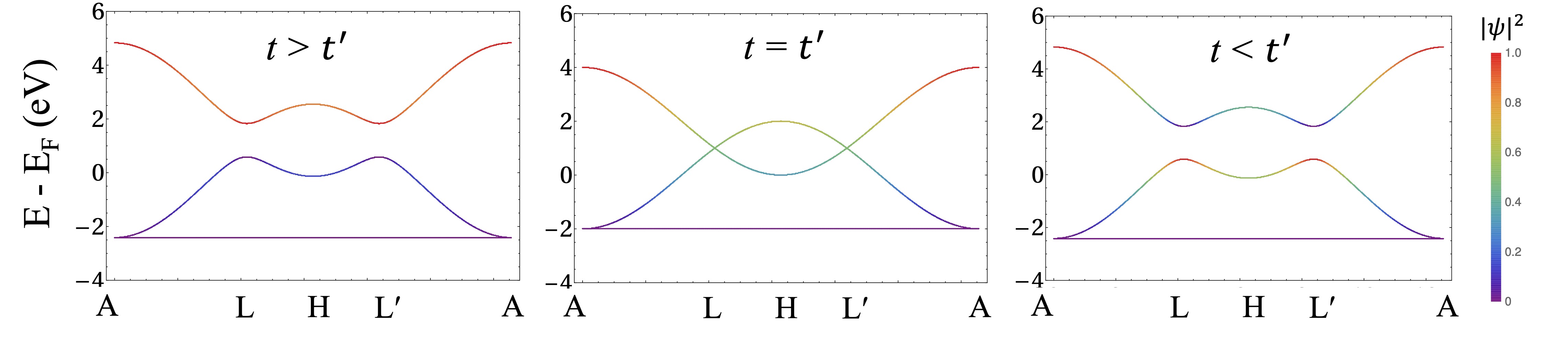}
\caption{\textbf{Band dispersions obtained from the TB model for different values of the intra-layer hopping parameters $t$ and $t^\prime$ in the $k_z = \pi$ planes.} The gap between the dispersive bands is trivial for $t > t^\prime$ (left). The gap closing critical point at the L is obtained for $t = t^\prime$ (middle). Further increase in $t^\prime$ reopens the gap with an orbital exchange at L, which signifies the non-triviality of the gap in the $k_z = \pi$ plane, shown in the right panel. }    \label{t_tp_bands}
\end{figure*}

As discussed in the main manuscript, for $k_z = \pi$ plane, the two layers in the coupled kagome lattice become decoupled, i.e., all the terms associated with inter-layer hopping parameter $t_v$ become zero. In other words, this means that for the $k_z = \pi$ plane, the dispersion follows the 2D kagome lattice. Here, we discuss the variation of band dispersion as a function of intra-layer hopping parameters $t$ and $t^\prime$. In Fig.~\ref{t_tp_bands}, we present the band dispersions for (left) $t > t^{\prime}$, (middle) $t = t^{\prime}$ and (right) $t < t^{\prime}$. The system is accompanied by a flat band irrespective of the values of $t$ and $t^{\prime}$. $t = t^{\prime}$ case resembles the dispersion for the ideal kagome system with a Dirac point at $L$. It can be observed that both for $t < t^{\prime}$ and  $t > t^{\prime}$ cases, bands forming Dirac point at $L$ become gapped. However, it can be noted that despite having similar bandstructure, these are not identical. The band inversion for $t < t^{\prime}$ makes this phase topologically non-trivial, which is distinct from the trivial phase appeared for $t > t^{\prime}$. In addition to the orbital exchange, as shown in Fig. ~\ref{t_tp_bands}, the non-triviality has further been confirmed by the Zak-phase calculation.

\clearpage

\end{document}